# Shift of $CO_2$-I absorption bands in diamond: a pressure or compositional effect? A FTIR mapping study

Evgenii P. Barannik[1], Andrey A. Shiryaev[2,*], Thomas Hainschwang[3]

1) Moscow Gemological Laboratory, Arbat St., 30/2, bld. 2, Moscow, Russia

2) A.N. Frumkin Institute of physical chemistry and electrochemistry, Russian Academy of Sciences, Leninsky pr. 31 korp. 4, 119071, Moscow

3) GGTL - GEMLAB Laboratory, Gnetsch, 42, LI – 9496 Balzers, Liechtenstein

* Corresponding author: Shiryaev A.A., email: a_shiryaev@mail.ru AND shiryaev@phyche.ac.ru

**Abstract**

Infra-red maps and profiles with high spatial resolution were obtained for two single crystal diamonds with pronounced $CO_2$ IR absorption peaks. Detailed examination allows unambiguous assignment of the spectral features to solid $CO_2$-I phase. It is shown that the distribution of IR band positions, intensities and widths in the sample follows regular patterns and is not chaotic as was suggested in previous works where spectra of a few individual spots were analysed. Consequently, pressure effects alone fail to explain all observed features and shifts of the $CO_2$ bands. Experimental data can be explained by presence of impurities (such as water, $N_2$, etc.) in the trapped $CO_2$. This implies that spectroscopic barometry of $CO_2$ microinclusions in diamond may be subject to poorly controlled bias. However, barometry is still possible if Davydov splitting of the $CO_2$-I $\nu_2$ band is unequivocally observed, as this indicates high purity of the $CO_2$ ice.

**Keywords:** diamond; $CO_2$ ice; Infra-red microscopy; barometry, impurity

# 1. Introduction

Carbon dioxide is an important metasomatic agent in Earth's mantle and plays a major role in the carbon cycle. The presence of $CO_2$ in diamonds was inferred long time ago from mass-spectrometry studies (Melton et al., 1972, Melton and Giardini, 1974, 1975, 1981). Presence of carbon dioxide in some gem quality single crystal diamonds follows from IR spectroscopy (Schrauder and Navon, 1993; Chinn, 1995) and cryomicroscopy of fluid inclusions (Voznyak et al., 1992; Tomilenko et al., 1997; Smith et al., 2014). Based on the pressure-induced shift of $CO_2$ peak positions Schrauder and Navon (1993) inferred extremely high residual pressures reaching 5 GPa in one specimen, suggesting the presence of compressed solid $CO_2$. Even higher residual pressures, up to 20 GPa, were mentioned in an extensive investigation of $CO_2$-containing diamonds by Chinn (1995). In that work strong variations in the shape of the $CO_2$-related bands between different samples and even between analysis points in the same specimen were reported.

Investigation of a large set of polished diamonds with $CO_2$ IR bands (termed as "$CO_2$ diamonds") using a beam condenser, i.e. allowing analysis of relatively small spots, revealed high variability of positions, widths, and relative intensities of $CO_2$ $v_3$ and $v_2$ bands (Hainschwang et al., 2006, 2008). These authors suggested that such a variability cannot be explained by the presence of $CO_2$ phase inclusions and tentatively proposed integration of $CO_2$-molecules in the diamond lattice. The presence of oxygen as a lattice impurity in diamond is plausible (Shiryaev et al., 2010 and references therein, Palyanov et al., 2016), but its correlation with the $CO_2$-absorption is uncertain. In this work, we report the results of a detailed investigation of diamond single crystals possessing $CO_2$ absorption bands using IR microscopy and discuss implications for diamond studies and for more general application of barometry of fluid inclusions in minerals based on spectroscopic data.

## 2. Samples and methods

Two $CO_2$-rich diamonds - FN7112 and FN7114 - were studied. Diamond FN7114 was described in Hainschwang et al. (2008). The samples were obtained commercially and their source is unknown. The stones were laser cut from gems and subsequently polished to make a double-sided plate. Both samples were treated at 6 GPa and 2100 °C for 10 minutes; the treatment did not influence the color and $CO_2$-related vibrations (Hainschwang et al., 2008). The sample FN7112 is 3.16 mm in diameter and 0.89 mm thick (mass 0.08 ct); the FN7114 dimensions are 2.51 mm and 1.03 mm (mass 0.06 ct), respectively. According to our previous X-ray topography study (Shiryaev et al., 2010), the samples are single crystals.

FTIR spectra were collected with a Nicolet iN10 FTIR microscope equipped with a liquid $N_2$-cooled MCT detector. The microscope and sample compartment were continuously purged with dry high-purity $N_2$ before and during spectral acquisition. Although strongly suppressed, traces of $CO_2$ gas were still observed in spectra. Assuming that this gas signal was from atmospheric contamination, a component of $CO_2$ gas (represented by the $\nu_2$ and $\nu_3$ bands) was included in the fit. The absorption of gaseous $CO_2$ for each spectrum was chosen in a way to minimize the difference between the experimental data and the model. This procedure is possible, firstly, because the shape of gaseous $CO_2$ absorption is known from independent atmospheric measurements, and secondly, because the bands of gaseous $CO_2$ and $CO_2$-I overlap only partially. All measurements were performed in transmission mode on free-standing samples at room temperature. For consistency, the microscope was focused on the upper surface of the sample. For the mapping, a square $150 \times 150$ μm$^2$ aperture and 100 μm step were used. For the profiles a $30 \times 30$ μm$^2$ aperture and 30 μm step were used. Due to high refractive index of diamond, the thickness-averaged size of the analyzing IR beam somewhat deviates from the quoted aperture size. Spectra were acquired in the 600-4000 cm$^{-1}$ spectral range. At least 64 scans per spectrum at a resolution of 2 cm$^{-1}$ were recorded.

FTIR spectra were processed using a custom Lua script for Fityk software (Wojdyr, 2010). The absorbance measurements were normalized to lattice absorption of a reference type IIa diamond sample; after the normalization, the spectrum of the type IIa reference was subtracted. Linear baseline correction was performed for each zone of interest. Most $CO_2$-related bands ($\nu_{2a}$, $\nu_{2b}$, $^{13}CO_2$ $\nu_{3b}$, $\nu_{3a}$) as well as carbonate and platelet bands were fitted with Gaussian distributions. Positions of relevant $CO_2$-related absorption bands are summarised in Table 1. The $CO_2$-I $\nu_{3b}$ band was approximated by two Gaussians with identical position. A Voigt profile was used for the peak at 3107 cm$^{-1}$ (tentatively assigned to a $VN_3H$-center, Goss et al., 2014). Concentrations of nitrogen in A and B form were calculated after spectral decomposition in the one-phonon region and using coefficients from Boyd et al. (1994) and Kiflawi et al. (1994). Tentative identification of minerals in inclusions was made using compilation by Chukanov and Chervonnyi (2016). Plots and maps were created using Matplotlib library for Python (Hunter, 2007). All raw and processed FTIR spectra are available as Supplementary Materials.

Table 1. Assignment of the $CO_2$-related bands (Osberg and Hornig, 1952; Lu and Hofmeister 1995).

| Position (cm$^{-1}$) | Band | Assignment |
|---|---|---|
| 650 | $\nu_{2b}$ | $CO_2$-I, bending ($\nu_2$) mode |
| 660 | $\nu_{2a}$ | $CO_2$-I, bending ($\nu_2$) mode |
| 2299-2305 | $^{13}CO_2$ $\nu_{3b}$ | $CO_2$-I, $^{13}CO_2$ analog of the $\nu_3$ mode |
| 2369-2375 | $\nu_{3b}$ | $CO_2$-I, symmetric stretching ($\nu_3$) mode |
| 2415 | $\nu_{3a}$ | uncertain |
| 3615 | $\nu_3 + 2\nu_2$ | $CO_2$-I, overtone |
| 3740 | $\nu_3 + \nu_1$ | $CO_2$-I, overtone |

**3. Results**

3.1. *The samples*

Both studied diamonds are single crystals with a light greenish-brown color. Diamond FN7112 shows faint yellow fluorescence under ultraviolet light excitation, while diamond FN7114 is inert. The diamonds contain tiny inclusions of roughly hexagonal shape with dimensions of 5-10 μm and thickness of less than a micron (Fig. 1). In sample FN7112 the inclusions are present in the whole body of the stone and are crystallographically oriented; in some regions larger plates appear to be surrounded by smaller ones, possibly indicating that decripitation of some inclusions has occurred. In sample FN7114 the inclusions occur in clusters of up to ~10 plates. It is important to emphasize that there is no obvious correlation between abundance and distribution of these inclusions and infra-red features (including $CO_2$-related bands) described below. Similar inclusions were reported by Lang et al. (2007) and Hainschwang et al. (2020) and ascribed to graphite. Raman microscopy of the inclusions (excitation 532 and 785 nm), X-ray diffraction, X-ray fluorescence and phase tomography have not provided useful information, thus unambiguous identification of the inclusions is not yet possible. However, the hypothesis of the presence of graphite or other types of $sp^2$-carbon in the inclusions now appears to be less sound, given the high sensitivity of Raman spectroscopy with visible excitation to $sp^2$-C and the absence of any relevant features in our spectra. In addition, needle-like inclusions with length up to ~40 μm are also present (see also Hainschwang et al. 2020). The later ones clearly do not correspond to the hexagonal plates viewed edge-on as suggested by careful examination under several inclination angles and generally larger sizes of the needles.

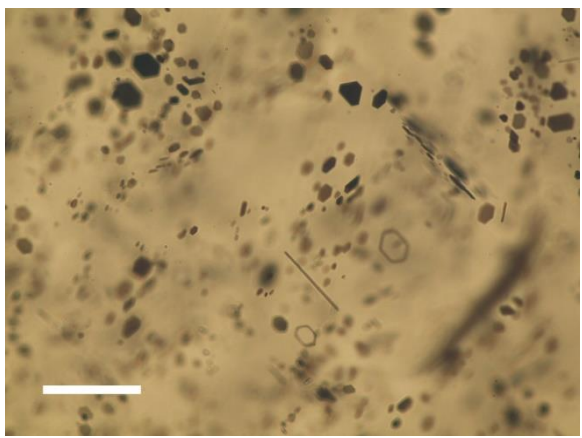

**Figure 1.** A photomicrograph of inclusions in diamond FN7112 in transmission light; scale bar – 50 μm. One can see numerous (quasi)hexagonal inclusions and rare needle-like ones (e.g., a ~40 μm-long one slightly below center of the image). Some of the inclusions do not contain dark material and at least in some cases they are very close to the surface of the stone, thus pointing to influence of the polishing on loss of the material (possibly due to heating).

In both samples unusual absorption features between ~900-1400 cm$^{-1}$ (one-phonon region) which cannot be ascribed to known nitrogen-related defects are observed (Fig. 2). Maxima of the bands are at 1060, 1120-1140, ~1245, ~1300 cm$^{-1}$, and a broad band with a maximum at ~1375 cm$^{-1}$ may also be related to this set of absorptions. A sharp peak at 1332 cm$^{-1}$ (Raman frequency of diamond) is always present, but its unique assignment is barely possible, since in crystals with diamond structure this vibration becomes IR-active in the presence of any defect, violating point symmetry of the matrix (Birman, 1974). These absorption bands were discussed in a previous study of $CO_2$-diamonds (Hainschwang et al., 2008). Similar features were observed in a recent study of diamonds from Chidliak, Canada (Lai et al., 2020); no microinclusions were observed microscopically in these samples (Lai, pers. comm). Broad bands, possibly related to water or OH-containing groups in inclusions, are observed in the 1450-1750 and 3200-3700 cm$^{-1}$ spectral regions for both samples (Fig. 2). Attempts to establish correlations between OH-related and carbonate bands were abandoned due to interference of broad bands with unknown origin, shapes and intensities.

Below we discuss in detail the spectral features of the studied diamonds.

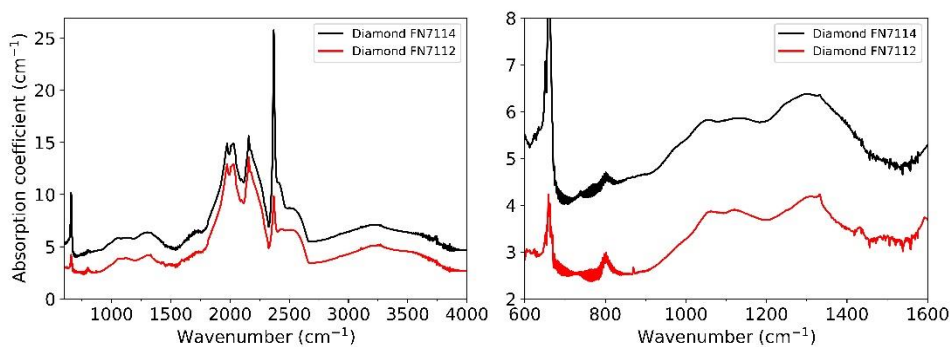

**Figure 2.** Typical FTIR spectra of the studied diamonds showing analysis points with $CO_2$-related absorptions. A – the whole recorded range, B – one-phonon region. For diamond FN7114, the spectrum of the brown zone is shown. For clarity, the spectrum of the specimen FN7114 is vertically displaced by 2 cm$^{-1}$.

*3.2. Diamond FN7112.*

A photomicrograph of the sample is shown in Fig. 3A. The FTIR spectra of specimen FN7112 show carbonate (870 and 1430 cm$^{-1}$) and 800 cm$^{-1}$ (possibly a silicate, e.g. quartz) bands, weak peaks at 3107 cm$^{-1}$ and 3272 cm$^{-1}$; maps of some of these features are shown in Fig. 3 B-D. Absorption by nitrogen-related A, B or C-centers is not detected. The bands at 650, 660 and 2370 cm$^{-1}$ are due to $CO_2$ $\nu_{2a}$, $\nu_{2b}$, and $\nu_{3b}$ vibrations (Fig. 4 A, B); weak features at ~3615 and ~3740 cm$^{-1}$ are $\nu_3 + 2\nu_2$ and $\nu_3 + \nu_1$ overtones, respectively (Fig. 4E). A very weak feature at 2300 cm$^{-1}$ is due to $^{13}CO_2$ $\nu_{3b}$. The bands show no rotational splitting and are shifted from the position of gaseous carbon dioxide. We assign these bands to the $CO_2$-I phase, see Discussion for detail. The most intense of the bands, the $\nu_{3b}$, appears symmetrical. In some spectra a shoulder at 2350 cm$^{-1}$ due to residual $CO_2$ gas in the spectrometer is visible.

The integrated area of the $CO_2$-I bands varies significantly across specimen FN7112 as shown by maps of $\nu_{2b}$, $\nu_{2a}$ and $\nu_{3b}$ peaks (Fig. 3 E-G). The $CO_2$ distribution is unrelated to strain observed in polarized light, but inversely correlates with intensity of the coloration: the $CO_2$ bands are stronger in less intensely colored zone. Apparently, the number of visible hexagonal inclusions is not correlated with the intensity of $CO_2$ absorption.

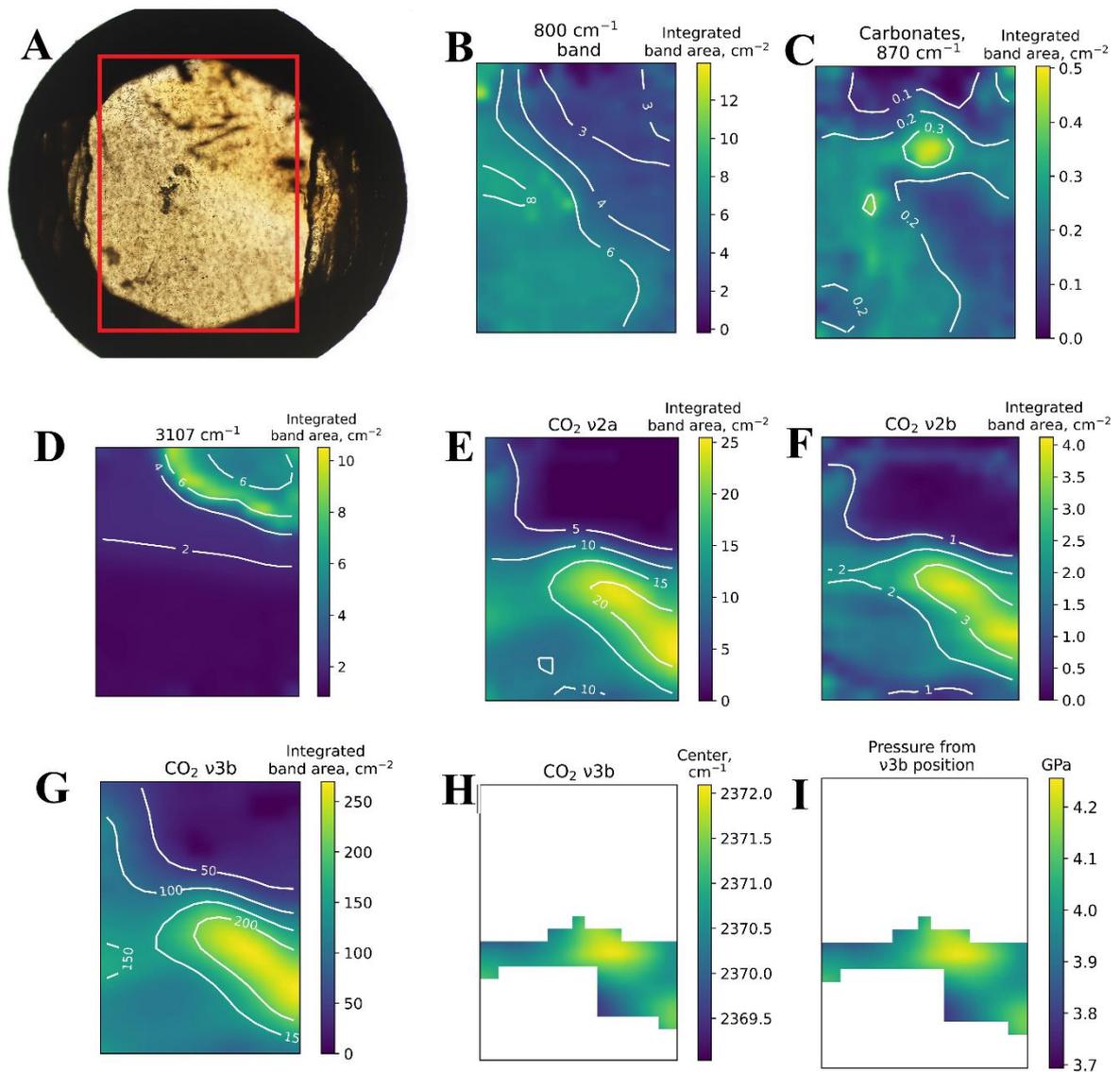

**Figure 3.** General view and FTIR maps of diamond FN7112. A - photomicrograph; red rectangle shows the position of the FTIR maps. Specimen diameter is 3.1 mm. Black rim is due to extinction in the crown; darkened roughly vertical fields to the right and left of the red rectangle – supporting polymer film. The FTIR maps are shown slightly larger than area marked on A to improve visibility. B – integrated area of 800 cm$^{-1}$ band, C – integrated area of carbonate band; D - Integrated area of 3107 cm$^{-1}$ peak. E-G – integrated areas of $CO_2$-related bands; H – position of $\nu_{3b}$ band; I – residual pressure in $CO_2$ inclusions calculated assuming absence of other factors influencing $\nu_{3b}$ band position (see text for detail).

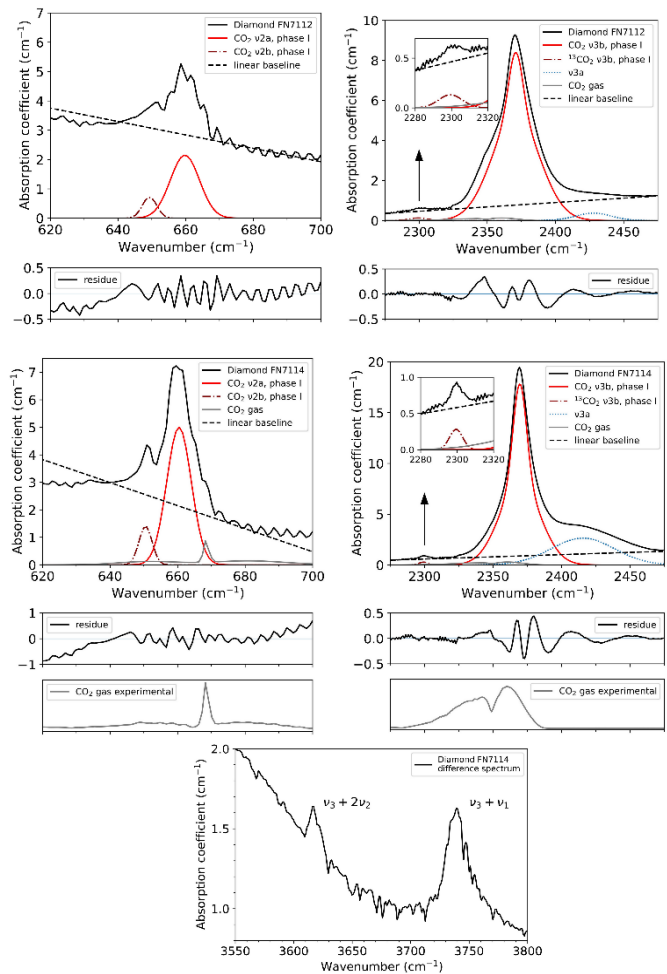

**Figure 4.** $CO_2$-related bands in FTIR spectra of the studied diamonds. A-D – examples of spectral decomposition for $\nu_2$ and $\nu_3$ regions. A, B – diamond FN7112; C, D – diamond FN7114. Experimental atmospheric spectra are shown for comparison. Residues show the difference between experimental data and the model. E - High frequency part of a diamond FN7114 spectrum.

*3.3. Diamond FN7114*

FN7114 is a diamond with distinct zoning; brown, yellow and (near) colorless zones are distinguished; IR-active features generally follow the zoning (Fig. 5). The colorless zone shows absorption by A, B-centers and platelets, superposed on unassigned bands (see Section 3.1). The three-phonon region possesses a complex shape between 3100-3300 cm$^{-1}$ with peaks at 3107, 3272 and very weak features at 3144, 3200, 3238, 3257 cm$^{-1}$. The ~800 cm$^{-1}$ band is rather uniformly distributed. Carbonates (possibly calcite) are detectable near the black feature visible in the right half of the diamond, which probably represents a healed fracture.

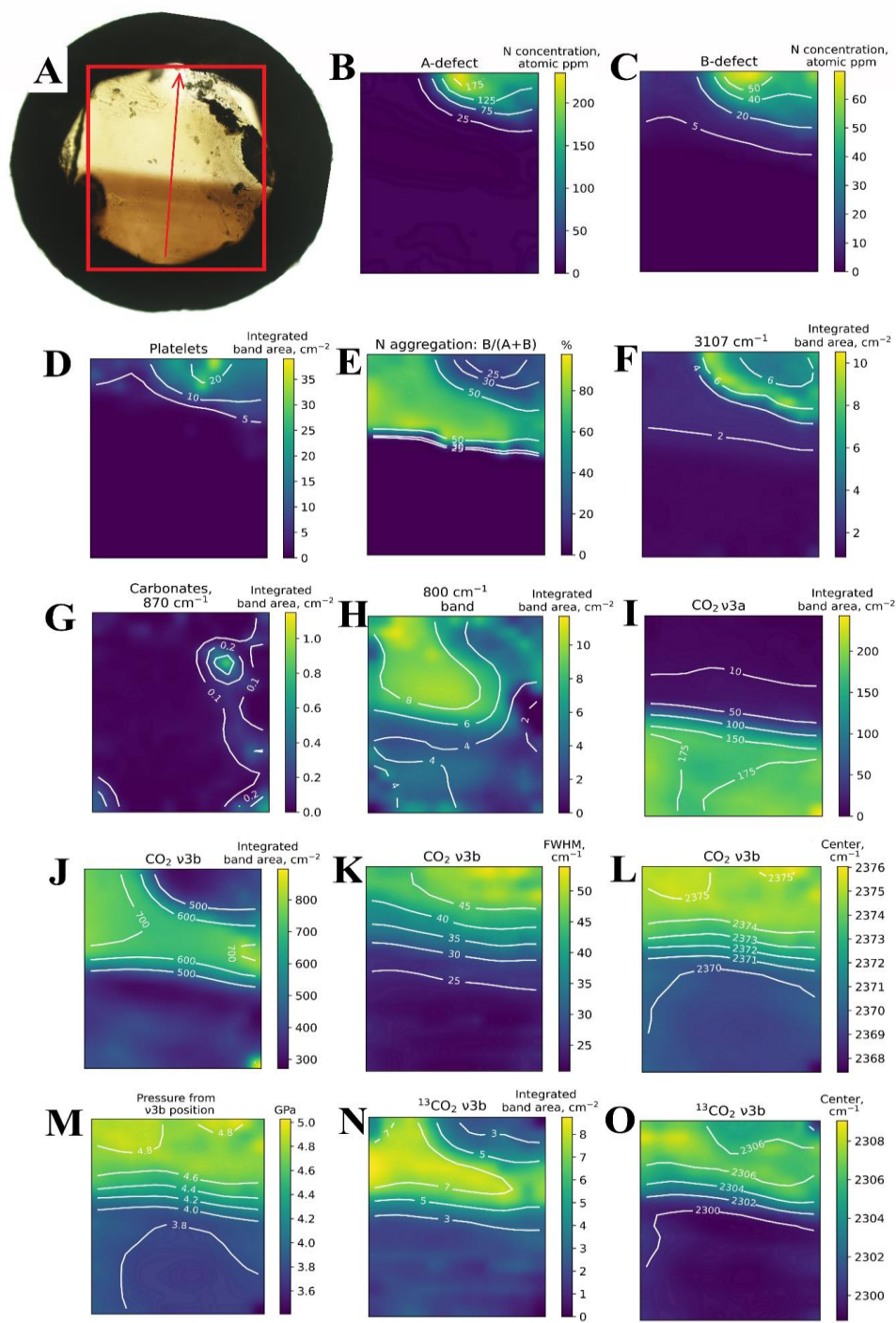

**Figure 5.** General view and FTIR maps of diamond FN7114. A - photomicrograph; red rectangle shows the position of the FTIR maps, arrow indicates direction of the profile. Specimen diameter is ~2.5 mm. Black rim is due to extinction in the crown; darkened roughly vertical fields to the left of the red rectangle – supporting polymer film. The FTIR maps are shown slightly larger than area marked on A to improve visibility. FTIR spectra of "black" mapped regions were obtained according to the standard procedure. FTIR maps: B, C, D, F – distribution of defects (see notations). E – nitrogen aggregation. Inclusions: G – integrated area of carbonate band; H – integrated area of 800 cm$^{-1}$ band. I-O – $CO_2$-related bands. I - integrated area of $v_{3a}$ band; J-L – integrated area, FWHM and position of $v_{3b}$ band. M – residual pressure in $CO_2$ inclusions calculated assuming absence of other factors influencing $v_{3b}$ band position (see text for detail). N, O – integrated area and position of $^{13}CO_2$ $v_{3b}$ band.

It is convenient to discuss changes of the spectral features in different zones of the specimen using profile across the sample shown by vertical arrow in Fig. 5A; evolution of various features is shown in Figure 6. Peaks due to carbon dioxide vary considerably in intensity and shape between the zones (Figs. 5, 6). The colorless zone shows weak $CO_2$-I bands; the yellow zone exhibits the strongest $CO_2$-I absorption and the brown zone spectra show a weak $CO_2$-I band with an additional feature at 2415 cm$^{-1}$ (Fig. 6). Assignment of the band at 2415 cm$^{-1}$ ($\nu_{3a}$) is uncertain, but several possibilities are examined in the Discussion section. In addition, a small band at 2300 cm$^{-1}$ is observed. We assign it to the $\nu_3$ vibration of isotopically-substituted carbon dioxide molecule, i.e. $^{13}CO_2$ $\nu_{3b}$; the supporting evidence is given in the Discussion section.

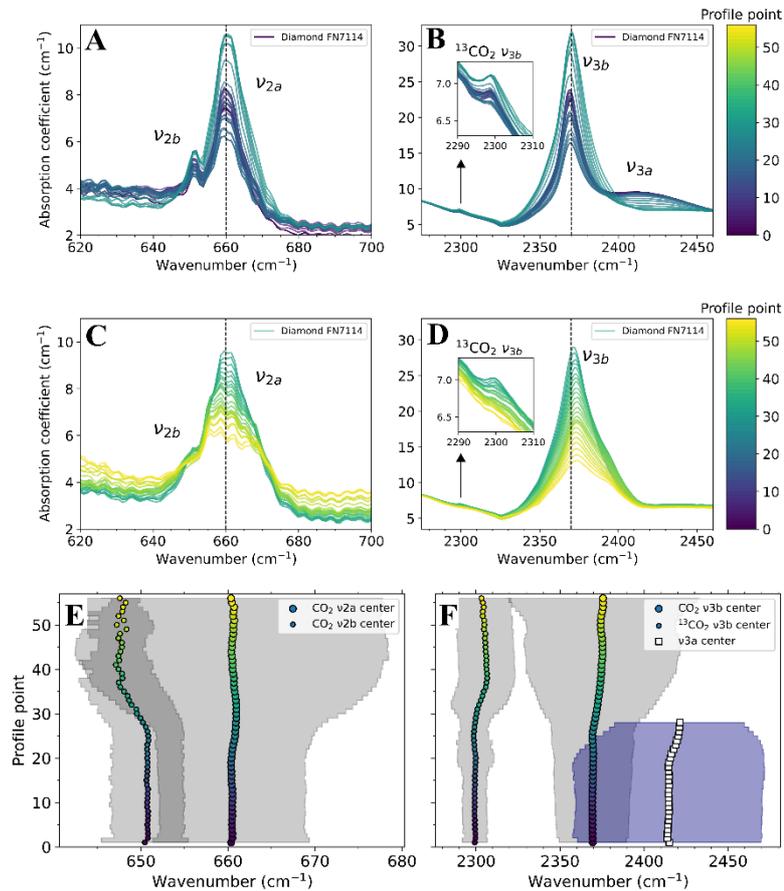

**Figure 6.** Evolution of $CO_2$-related bands in FTIR spectra along the profile across diamond FN7114. For A-D spectra are normalized, but no other manipulations are performed. A, B – spectra of the brown zone. C, D – spectra of the yellow and colorless zone. Dashed vertical lines at 660 (A, C) and 2370 cm$^{-1}$ (B, D) are provided for visual reference. E, F – changes in band position and FWHM. Markers show band center positions; shaded regions are set to ±FWHM. The color scheme shows profile point number.

Along the profile, the CO$_2$-I $\nu_{3b}$ band broadens (FWHM increases from 20 to 55 cm$^{-1}$) and its center shifts from 2369 to 2375 cm$^{-1}$; the $^{13}$CO$_2$ $\nu_{3b}$ isotopic band behaves in a similar way (Fig. 6). The position of the $\nu_{2a}$ band does not change significantly, while its FWHM increases from 8 to 17 cm$^{-1}$. Upon broadening of the more intense $\nu_{2a}$ band, the $\nu_{2b}$ component turns into a shoulder, introducing uncertainty into its position for the zone containing nitrogen defects.

Examination of distribution of the CO$_2$-related and carbonates bands does not allow to establish a solid correlation. Whereas for diamond FN7112 these phases might be correlated, for sample FN7114 no relation is apparent. Therefore, it is unclear whether CO$_2$-carbonate correlation exists.

Figure 7A shows evolution of integrated areas of the CO$_2$-related bands and nitrogen concentration along the profile. At the transition from the brown to yellow zone (points 20-30), the absorption of CO$_2$-I rapidly increases and the $\nu_{3a}$ band disappears. Simultaneously, defect-related bands in the 900-1400 cm$^{-1}$ region undergo a complex shape change; total absorption diminishes; the peak at 1332 cm$^{-1}$ becomes more prominent. In the same time, absorption in the ranges 1450-1750 and 3200-3700 cm$^{-1}$ becomes stronger.

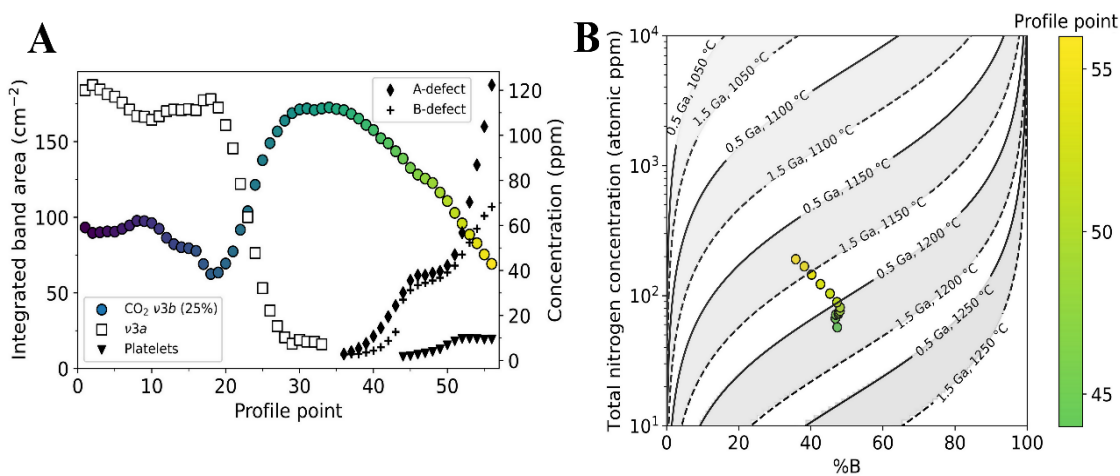

**Figure 7.** Evolution of CO$_2$ and N-related IR features along the profile across diamond FN7114. A — concentration of nitrogen-related defects and areas of $\nu_{3a}$ and CO$_2$-I $\nu_{3b}$ (the area of CO$_2$-I $\nu_{3b}$ band is multiplied by 0.25). B - Nitrogen aggregation plot for diamond FN7114. The activation energy and Arrhenius constant are from Taylor et al. (1990). The data are corrected for laboratory-based HPHT annealing described by Hainschwang et al. (2008). The color scheme shows profile point number.

Since the studied sample represents a laser-cut polished piece of a gem diamond, identification of the growth direction is not trivial. X-ray topography does not reveal obvious growth dislocations, partly due to the moderate degree of plastic deformation. However, examination of a plot of nitrogen concentration vs N aggregation state (Fig. 7B) might give a clue. If one assumes that the diamond growth proceeded in a gradually cooling system, the N aggregation plot suggests that the N-containing zone reflects late stages of the crystal formation. This does not necessarily imply absence of N in the growth medium; it rather indicates unfavorable conditions for incorporation of this impurity. The content of N in the form of common A and B defects gradually increases towards the end of the profile (Fig. 7A). We note, however, that any hypothesis about the N concentration in the FN7114 diamond may be somewhat simplistic, since we do not know yet whether the defects giving rise to absorbance in the one-phonon region contain nitrogen or not.

## 4. Discussion

*4.1. Band assignment.*

*4.1.1. Major $CO_2$-I bands*

At room temperature, $CO_2$ crystallizes into a cubic phase I between 0.6 GPa and 2 GPa (Lu and Hofmeister 1995, Olijnyk and Jephcoat, 1995); the variations in the transition pressure are ascribed to the size of the $CO_2$ droplets and, possibly, the nature of the surrounding medium. In the pressure range of ~8-13 GPa a transition into orthorhombic $CO_2$-III occurs (Lu and Hofmeister, 1995). Fundamental modes and, consequently, IR spectra of these phases differ. The IR spectrum of $CO_2$-I shows two bending bands ($v_{2a}$, $v_{2b}$) and a single asymmetric stretching band ($v_3$). The $CO_2$-III spectrum is characterized by splitting of the fundamental modes, showing in total three bending and two stretching bands. With increasing pressure, the stretching and bending vibrations have positive and negative shifts, respectively, in both phases I and III

(Hanson and Jones, 1981; Lu and Hofmeister, 1995). If the $\nu_2$ and $\nu_3$ bands observed in our work indeed arise from the $CO_2$–I phase, a linear correlation between their areas is expected. Figure 8 shows correctness of this assumption for both samples. We emphasize that the spectra of the brown zone in diamond FN7114 show two main components in the $\nu_3$ region — $\nu_{3a}$ and $\nu_{3b}$. Only $\nu_{3b}$ appears to correspond to the $CO_2$-I phase; areas of $\nu_{3a}$ and $CO_2$-I bands are inversely correlated. Based on the observed number of bands, positions and $\nu_{2b}+\nu_{2a}$ to $\nu_{3b}$ area correlation, we can confidently assign $\nu_{2b}$, $\nu_{2a}$, $\nu_{3b}$ to the $CO_2$-I phase.

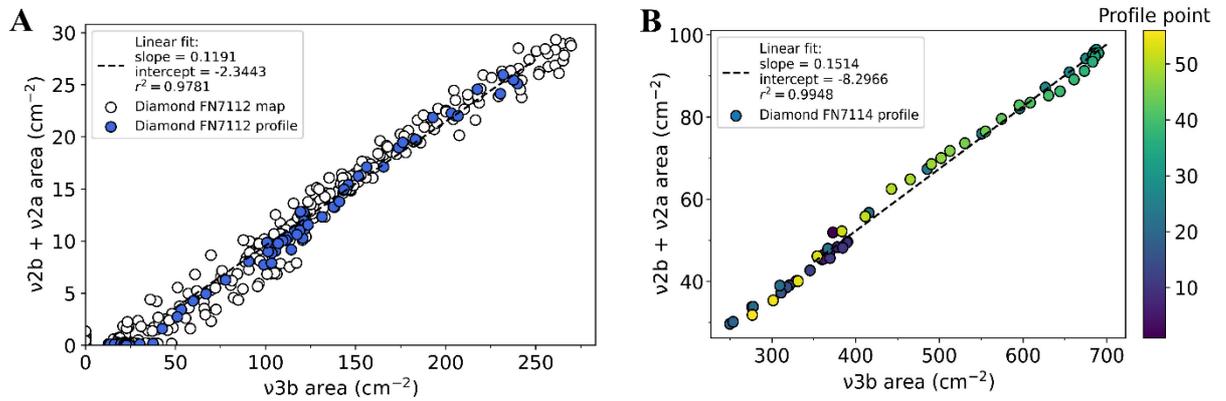

**Figure 8.** Correlation of the integrated area of $CO_2$ $\nu_2$ and $\nu_3$ bands for studied diamonds. A – $\nu_{2a}+\nu_{2b}$ and $\nu_{3b}$ bands for diamond FN7112. Datasets for both map and profile through diamond FN7112 are shown. B – $\nu_{2a}+\nu_{2b}$ and $\nu_{3b}$ bands for diamond FN7114.

A $\nu_{2b}$ band centred at 650 cm$^{-1}$ contributes up to 15% to the $\nu_{2a}$ (660 cm$^{-1}$) area. It behaves differently from that of $\nu_{3b}$ and the 2415 cm$^{-1}$ peak (the $\nu_{3a}$ band, see below), transforming into a shoulder in the nitrogen-rich zone of the diamond. The $\nu_{2a}$ and $\nu_{2b}$ bands correspond to $CO_2$ bending influenced by Davydov splitting. Lattice disorder markedly influences the magnitude of the Davydov splitting in molecular crystals. In particular, addition of 10% of $H_2O$ into $CO_2$ ice leads to suppression of the splitting (Baratta and Palumbo, 2017). Consequently, the presence of distinct $\nu_{2b}$ absorption may indicate purity of $CO_2$-ice in inclusions.

*4.1.2. Isotopic $^{13}CO_2$-I $v_{3b}$ band*

A weak band at 2300 cm$^{-1}$ (Fig. 4 B,D) is observed in the whole mapped region of sample FN7114 (Fig. 5K) and in zones with relatively high $CO_2$ absorption in diamond FN7112 (Fig. 4B). The measurement of the 2300 cm$^{-1}$ band is complicated due to its superposition on intrinsic diamond absorption bands. For the profile across the sample FN7114, the area of 2300 cm$^{-1}$ band was measured to be 0.5-1.2% of $v_{3b}$ area. This band is redshifted 70 cm$^{-1}$ from the $CO_2$ $v_{3b}$ peak and its position varies along the profile from 2299 to 2305 cm$^{-1}$ whereas the $v_{3b}$ peak shifts from 2369 to 2375 cm$^{-1}$. The FWHM and spatial distribution of the $v_{3b}$ and 2300 cm$^{-1}$ bands behave in a similar manner. In harmonic approximation, substitution with a heavy isotope should redshift absorption bands and the area of the relevant band should correspond to the isotopic fraction, i.e. ~1.1 % in the case of carbon. For crystalline $CO_2$ the $v_3$ $^{13}CO_2$ band is located at 2283 cm$^{-1}$ and is redshifted 62 cm$^{-1}$ from $v_3$ $^{12}CO_2$ peak at 2345 cm$^{-1}$ (Dows and Schettino, 1973). Based on the observed changes in intensity, position and width, the 2300 cm$^{-1}$ band can be assigned to the $v_{3b}$ mode of $^{13}CO_2$; thus we denote the 2300 cm$^{-1}$ band as $^{13}CO_2$ $v_{3b}$.

*4.1.3. The 2415 cm$^{-1}$ ($v_{3a}$) band*

A band at 2415 cm$^{-1}$ denoted as $v_{3a}$ (Chinn, 1995) is observed in spectra of FN7114, but is practically absent in spectra of diamond FN7112 (Fig. 4 B,D). This band can be enhanced by HPHT treatment of $CO_2$ diamonds (Hainschwang et al. 2008). According to our data, the $v_{3a}$ band area can reach 70% of the $v_{3b}$. Analyzing its behavior along the profile, one can see that this band is relatively intense in the brown part of the sample FN7114, but almost disappears closer to the N-containing zone. The gradual disappearance of this band correlates with an increase of broad features between 3200-3700 and 1450-1750 cm$^{-1}$ (possibly, OH-related) and a weak peak at 800 cm$^{-1}$. Judging from spectra with prominent 2415 cm$^{-1}$ band, its position and FWHM (~55 cm$^{-1}$) are approximately constant. The assignment of this band is uncertain, but several possibilities are discussed below.

Hainschwang et al. (2008) assigned the $\nu_{3a}$ band to a highly shifted $\nu_3$. However, detailed examination shows that the $\nu_{3a}$ band cannot be assigned to the $CO_2$-I or $CO_2$-III macroscopic phases. The value of 2415 cm$^{-1}$ exceeds wavenumbers possible for $CO_2$-I phase as it would imply pressures outside of the $CO_2$-I stability field. For the $CO_2$-III phase an increase of pressure from 9 to 20 GPa shifts splits $\nu_3$ bands from 2408 to 2436 cm$^{-1}$ and from 2351 to 2369 cm$^{-1}$ (Lu and Hofmeister, 1995). The assignment of the 2415 cm$^{-1}$ band to the $CO_2$-III phase would also imply existence of a more intense feature centered at ~2355 cm$^{-1}$, which is actually absent. Secondly, the $CO_2$ bending mode in spectra from sample FN7114 is doubly split, showing only $\nu_{2a}$ and $\nu_{2b}$ components, excluding the $CO_2$-III phase. However, the $\nu_{3b}$ and $\nu_{3a}$ bands appear to be closely linked. Figure 7A shows that areas of $\nu_{3a}$ and $\nu_{3b}$ bands are inversely related by a factor of 2-3. Moreover, if the $\nu_{3a}$ band is assigned to a shifted $\nu_3$, the linear dependence between the $\nu_2$ and $\nu_3$ area is violated for sample FN7114 (Fig. 9). If we assume that the $\nu_{3a}$ band is not related to the $\nu_3$ mode of $CO_2$-I, the linearity is restored.

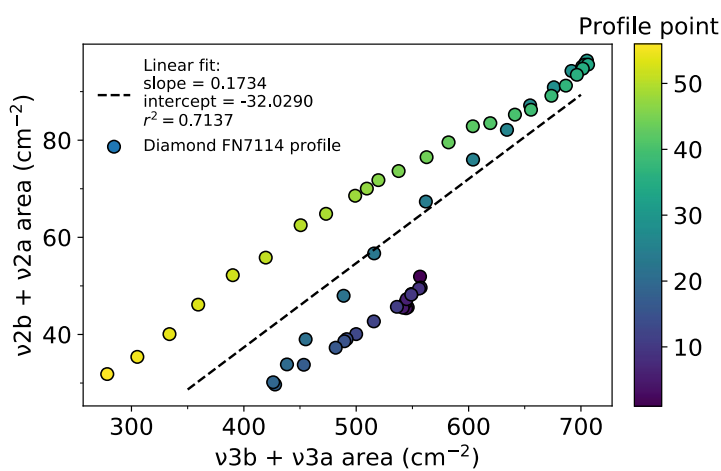

**Figure 9.** Correlation of the integrated areas of $\nu_{2a}+\nu_{2b}$ and $\nu_{3b}+\nu_{3a}$ bands for diamond FN7114. Presence of several segments is explained by evolution of areas of corresponding bands along the profile. Profile points between 1-20 and 20-35 are influenced by $\nu_{3a}$ and thus deviate from linearity.

The 2415 cm$^{-1}$ band could be a $\nu_3$ LO (longitudinal optical) phonon mode, provided that the thickness of the substance is sufficiently small and/or the incident IR beam is not perpendicular to the $CO_2$ film surface (Berreman, 1963). At low temperatures in pure $CO_2$ ice the LO phonon is blueshifted 39 cm$^{-1}$ from the TO (transverse optical) mode. The shape and maximum position of the LO band depends on the presence of impurities in the ice (Cooke et al., 2016). It should be noted, however, that we do not observe features unambiguously assigned to the $\nu_2$ LO phonon mode.

And last but not least, the $\nu_{3a}$ feature may represent hydrogen bonds in hydrogen carbonates. For example, bands with important features in the relevant spectral range are observed in $CO_2 \cdot H_2O$ complex and carbonic acid (Zheng and Kaiser, 2007); broadly similar spectra are reported for $KHCO_3$ single crystal (Lucazeau and Novak, 1973). Albeit it readily decomposes at ambient conditions, carbonic acid is stable at pressures of several GPa and rather high temperatures (Wang et al., 2016, Abramson et al., 2017). Of course, IR spectra of both carbonic acid and $KHCO_3$ contain numerous bands absent in our case. Nevertheless, hydrogen bonds in protonated carbonate species are among plausible species responsible for the 2415 cm$^{-1}$ peak, since local environment of $HCO_3^-$ ions may lead to marked spectral variations. If this assumption is correct, behavior of this band along the profile may correspond to gradual disappearance of the protonated compounds into $CO_2$ and aqueous solution. A related process of diamond formation by pH drop was recently proposed by Sverjensky and Huang (2015).

*4.2. Shift of $CO_2$ IR bands*

Several independent mechanisms may be responsible for blueshift of $CO_2$ $\nu_3$ absorption bands and may influence their shapes: a) residual pressure; b) size, shape and eventual heterogeneous structure of $CO_2$ precipitates; c) strong interaction of $CO_2$ with matrix; d) presence of impurities in the $CO_2$. These possibilities are considered in detail below.

*4.2.1. Pressure effects.*

In the case of $CO_2$-containing microinclusions in diamond, the most common explanation for the shifted $CO_2$ absorption bands is high residual pressure (e.g., Schrauder and Navon, 1993,

Chinn, 1995, Shiryaev et al., 2006). Residual $CO_2$-I pressure may be estimated from the pressure shift of the absorption bands using the experimental data of Hofmeister and Lu (1995) and Hanson and Jones (1981). In our work spectra with a $CO_2$ $\nu_{3b}$ area of at least 125 cm$^{-2}$, for which relatively precise peak position measurements were possible, were used for this purpose. The calculated pressures vary significantly for both samples. In sample FN7114 the values are 3.7-5.0 GPa (Fig. 5O). The observed positions of $\nu_{2a}$ and $\nu_{2b}$ also fit the expected values for ~4 GPa. However, the $\nu_{2a}$ pressure derivative is too small for meaningful calculations using the available data; precise determination of $\nu_{2b}$ position is difficult due to its transformation into a shoulder. For sample FN7112 only some points have $\nu_{3b}$ area of more than 125 cm$^{-2}$, so reported maps are incomplete. For sample FN7112 calculated pressures are 3.7-4.2 GPa (Fig. 3I).

The principal problem with the attribution of the blueshifted bands to pressure effects only lies in the peculiar spatial distribution of inferred pressures in the diamonds. This was noted already in previous works (Schrauder and Navon 1993, Chinn 1995, Hainschwang et al. 2008) and is addressed in more detail in the present work. In sample FN7112, the $CO_2$ bands with the largest shift (the highest inferred pressure) form irregular zones. In diamond FN7114, the inferred residual pressure increases upon approaching the nitrogen-containing zone. The nitrogen aggregation plot for the latter diamond indicates that temperature was most likely decreasing during crystallization of the N-containing zone, which is difficult to reconcile with increasing pressure. Note, however, that nitrogen may be contained in IR-inactive defects or in unidentified defects with unknown IR cross-section. The IR beam may eventually average over several growth horizons, although this contribution appears to be minor, since the trend on the Taylor plot does not show sharp or erratic changes.

From calculations based on elasticity theory, Anthony and Meng (1997) suggested that certain spatial distributions of microinclusions may indeed generate stresses approaching the crushing strength of diamond and lead to plastic deformation. In principle, partial release of the residual pressure by the plastic flow might be able to explain the gradual changes of $CO_2$ bands positions. Chinn (1995) suggested that extremely high pressures in some of diamonds are explained by partial graphitization of the inclusion walls, see Anthony (1994) for relevant calculations. However, several observations indicate that pressure alone cannot be the cause of the band behavior: a) the pressure hypothesis does not explain the observed decrease of the absorption and band broadening; b) lack of correlation between the inclusions observed microscopically and the $CO_2$ absorption peaks casts doubt on the "stress" hypothesis. X-ray topographs of these stones shows numerous small specks, which are almost certainly due to localized strain fields surrounding hexagonal inclusions. In the same time, X-ray topography, being a sensitive method, shows rather moderate degree of deformation and distribution of stress in the samples does not follow evolution of the $CO_2$-related IR bands.

*4.2.2. Crystallography and structure of $CO_2$ precipitates*

In thin crystals with cubic symmetry a LO phonon may appear in transmission mode IR measurements (Berreman, 1963); for thin layers of $CO_2$ ices this was well demonstrated by Escribano et al. (2013). This effect may strongly distort the "standard" spectrum and novel peaks may appear as blueshifted bands of $CO_2$. It was also shown (Signorell et al., 2006; Isenor et al., 2013) that in core-shell $CO_2$-$N_2O$ and $CO_2$-$H_2O$ nanoparticles the spectral envelope very strongly depends on the thickness of the layers (particle structure), shape and composition of the particles. Some of the $CO_2$-diamonds from Chinn (1995) (e.g., GC 859C, 727H, 874-C1 and C-2, 790B) show spectra remarkably similar to those modelled and measured by Isenor et al. (2013). Variations of composition and structure of the $CO_2$-containing inclusions may explain at least part of the observed complexity. We also note that, due to limited spectral resolution (2 cm$^{-1}$) of our experiment, variations in chemistry, shape, and stress state experienced by the $CO_2$-species, fine structure of the $CO_2$ absorption bands may be unresolved.

*4.2.3. Matrix interaction*

Hainschwang et al. (2008) used the anomalous shifts, FWHM and intensities of bands assigned to $CO_2$ to propose that in the studied diamonds those features cannot be explained by microinclusions of $CO_2$. Instead, they suggest that exsolution of oxygen impurities in the diamonds may form $CO_2$ molecules and that their interaction with the diamond lattice is responsible for the observed spectral peculiarities. This scenario is qualitatively similar to the behaviour of $CO_2$-molecules in channels present in the structure of cordierite and some other minerals (see Chukanov and Chervonnyi (2016) and references therein). For the diamond case, models resembling sub-nm voids with fullerene-like walls were considered, but correspondence between the calculations and experimental spectra is, at best, qualitative (Adjizian et al., 2009). Refinement of this model requires very detailed analysis of atomic structure of the void walls, not performed yet. Assignment of the observed *regular* evolution of IR bands across diamond crystals to the voids filled with pure $CO_2$ necessitates a mechanism of similar changes in the void structure.

*4.2.4. Influence of impurities in $CO_2$.*

A hypothesis, which, in our view, better fits experimental data, is discussed below. The presence of impurities in $CO_2$ ice may influence positions, width and splitting of IR features. For example, in mixed $CO_2$-$H_2O$ ice an increase of water fraction from 22 to 75% leads to blueshift of the $\nu_3$ maximum by almost 9 cm$^{-1}$ (Cooke et al., 2016). Judging solely by the position of the $\nu_3$ maximum, such blueshift can be interpreted as an increase in pressure by ~1.5 GPa, which is obviously not the case. As already mentioned, admixture of water also leads to band broadening and a loss of $\nu_2$ Davydov splitting occurs (Cooke et al., 2016). Raman study of fluid $N_2$-$CO_2$ inclusions in natural diamonds also suggest a strong influence of chemical composition on spectral features (Smith et al., 2014). Examination of a profile across the FN7114 sample shows gradual evolution of position, FWHM and intensity of the $CO_2$-bands. In the nitrogen-rich zone

$CO_2$-I bands become weaker, broader and shifted. All those phenomena can be explained by an increase of the impurity fraction in the $CO_2$ ice.

*4.3. Implications for oxygen in diamond.*

Conclusive evidence for oxygen impurities in the diamond lattice remains elusive. However, there is a growing body of evidence that several spectroscopic features of diamonds could be related to this impurity: an optical absorption band with a peak at 480 nm (Hainschwang et al., 2008), a peak at 1060 cm$^{-1}$ in IR (Malogolovets and Nikityuk, 1978, Palyanov et al., 2016) and the 566 nm luminescence band (Palyanov et al., 2016). Earlier works (Hainschwang et al 2008, Schrauder and Navon, 1993) indicated that the $CO_2$ IR absorption is stronger in regions with lower concentrations of N-related defects, see also Fig. 7 A from the present study. SIMS investigation of a large set of diamonds showed the existence of a positive correlation between total N and O contents (Shiryaev et al., 2010). Since the later element is clearly present as a substitutional impurity, this correlation appears to support the hypothesis of oxygen-related defects in the diamond lattice. We stress that the proportionality coefficient between N and O in SIMS data for the $CO_2$ and pseudo-$CO_2$ diamonds differs considerably from those in samples lacking $CO_2$ IR features. It was suggested that the coexistence of these two impurities in a given volume and their interaction prevents formation and/or hinders IR manifestations of well-known N defects and at least part of the SIMS and X-ray scattering results can be explained by formation of N-O-containing inclusions.

One of the main results of the present study is that the position and intensity of the $CO_2$ peaks in diamonds are not entirely random, but, instead, gradual changes and/or domains enriched in $CO_2$ are present. In many cases, crystallographic zoning revealed by cathodoluminescence is present (Chinn, 1995, Hainschwang et al., 2008). All studies of $CO_2$-diamonds mention anticorrelation of the carbon dioxide features and IR-active N and H impurities in the diamond lattice. Our detailed FTIR mapping and profiling show that although some amount of structural hydrogen and nitrogen is definitely present in all parts of the studied samples as manifested by the 3107 cm$^{-1}$ peak, the $CO_2$-absorption bands indeed become weaker with increased concentration of common A and B defects. Interestingly, diamond synthesis in $CO_2$-rich alkaline systems always produces N-rich crystals (Khokhryakov et al., 2016, Palyanov et al., 2016). Incorporation of nitrogen in diamond is clearly a complex function of N speciation and abundance in the growth medium, which, in turn, depend on its composition and $fO_2$. Therefore, the $CO_2$-N(H) dependencies suggest the importance of the composition of the growth medium on the prominence of the $CO_2$-features. Changes of fluid chemical composition will be pronounced not only in IR spectra of hydrous fluids or melts captured by diamonds (Zedgenizov et al., 2005; Weiss et al. 2013), but also in spectra of $CO_2$-bearing diamonds.

## 5. Conclusions

A detailed FTIR investigation of polished plates cut from two $CO_2$-rich single crystal diamonds reveals that changes of the $CO_2$-related IR features are not as random as it might seem from examination of bulk samples. Regular evolution of the position of the bands, FWHM and intensity is recorded for one of the samples; for the second stone, several domains with highly variable spectra are observed. Consideration of various possible mechanisms responsible for blueshift of $CO_2$-absorption bands suggests that the observed spectral shifts cannot be explained exclusively by the residual pressure assumption. In our view, accounting for impurities (primarily aqueous and N-containing species) in entrapped $CO_2$ ice is necessary for consistent explanation of the data. In future works spatially resolved spectroscopic measurements are preferred over the bulk ones, otherwise strong peak overlap may occur.

An important implication of our results is that shifts of $CO_2$-related bands in diamonds should be employed as a barometer with great care. If unaccounted for, impurities in $CO_2$ ice can introduce significant bias. In high purity $CO_2$-I the $\nu_2$ band is subject to Davydov splitting. Consequently, $CO_2$ spectroscopic barometry gives unambiguous results only in cases when

Davydov splitting of $CO_2$ $\nu_2$ band is clearly observed. Schrauder and Navon (1993) reported a strong asymmetry of the $\nu_2$ $CO_2$ band, but in their spectrum there is no obvious Davydov splitting. This may indicate the presence of impurities in $CO_2$ ice in their sample. Thus, the reported residual pressure appears to be overestimated as it was calculated solely based on the position of the $CO_2$ bands.

Our results allow to explain the nature of $CO_2$ related bands in $CO_2$-diamonds by the presence of impure $CO_2$-I in microinclusions. However, this does not exclude the possibility of the presence of oxygen as a lattice impurity in diamonds.

**Supplementary materials**

All raw and processed FTIR spectra are available as Supplementary Materials at: "FTIR maps and profiles for CO2-rich diamonds", Mendeley Data, V2, doi: 10.17632/yjrgv5fhhm.2.

**Acknowledgments**

The study was partly supported by RFBR grant 13-05-91320-SIG-a to AAS. We thank Dr. A. Shapagin for access to FTIR microscope and Uladzislava Dabranskaya from Utrecht University for creation of the graphical abstract. We highly appreciate thorough consideration of the manuscript and highly useful comments made by two anonymous reviewers.

**References**

Abramson E.H., Bollengier O., Brown J.M. (2017) Water-carbon dioxide solid phase equilibria at pressures above 4 GPa, *Scientific Reports*, 7, 821. DOI:10.1038/s41598-017-00915-0

Adjizian J-J., Latham C.D., Heggie M.I., Briddon P.R., Rayson M.J., Öberg S. (2009) Infra-red signatures of $CO_2$ in diamond from first principles. Ext. Abstracts, 60th De Beers Diamond Conference.

Anthony T.R. (1994) The behavior of gas inclusions in diamond generated by temperature changes, *Diam. Relat. Mater.*, 4, 83-94.

Anthony T.R. and Meng Y. (1997) Stresses generated by inhomogeneous distributions of inclusions in diamonds, *Diam. Relat. Mater.,* 6, 120-129.

Baratta G.A., Palumbo M. E. (2017) The profile of the bending mode band in solid $CO_2$. *Astronomy & Astrophysics*, 608, A81 (9 pp).

Berreman D.W. (1963) Infrared Absorption at Longitudinal Optic Frequency in Cubic Crystal Films, *Phys. Rev*., 130(6), 2193-2198.

Birman J.L. (1974) Theory of Crystal Space Groups and Infra-Red and Raman Lattice Processes of Insulating Crystals. In: Theory of Crystal Space Groups and Lattice Dynamics. Springer, Berlin, Heidelberg. https://doi.org/10.1007/978-3-642-69707-4_1

Boyd S.R., Kiflawi I., Woods G.S. (1995) Infrared absorption by the B nitrogen aggregate in diamond, *Philosophical Magazine B*, 72(3), 351-361.

Chinn I. (1995) A Study of Unusual Diamonds from the George Creek K1 Kimberlite Dyke, Colorado. PhD Thesis, University of Cape Town.

Chukanov N.V., Chervonnyi A.D. (2016) Infrared Spectroscopy of Minerals and Related Compounds, Springer.

Cooke I.R., Fayolle, E.C., Öberg, K.I. (2016) $CO_2$ Infrared Phonon Modes in Interstellar Ice Mixtures. *Astrophys. J.*, 832 (1), 5 (8 pp). https://doi.org/10.3847/0004-637X/832/1/5.

Dows D.A., Schettino V. (1973) Two-phonon infrared absorption spectra in crystalline carbon dioxide, *J. Phys. Chem.,* 58, 5009-5016; doi: 10.1063/1.1679088

Escribano R.M., Muñoz Caro G.M., Cruz-Diaz G.A., Rodríguez-Lazcano Y., and Maté B. (2013) Crystallization of $CO_2$ ice and the absence of amorphous $CO_2$ ice in space. *Proc. Nat. Acad. Sci,* 110(32), 12899–12904.

Goss J.P., Briddon P.R., Hill V., Jones R., Rayson M.J. (2014) Identification of the structure of the 3107 cm$^{-1}$ H-related defect in diamond. *J. Phys.: Condens. Matter*, 26(14), 145801. doi:10.1088/0953-8984/26/14/145801

Hainschwang T., Franck N., Emmanuel F., Laurent M., Christopher M.B., Rondeau B. (2006) Natural "$CO_2$-Rich" Colored Diamonds. *Gems & Gemology*, Fall 2006, 97.

Hainschwang T., Notari F., Fritsch E., Massi L., Rondeau B., Breeding C.M., Vollstaedt H. (2008) HPHT Treatment of $CO_2$ Containing and $CO_2$-Related Brown Diamonds. *Diam. Relat. Mater.*, 17(3), 340–351. https://doi.org/10.1016/j.diamond.2008.01.022

Hainschwang T, Notari F., Pamies G. (2020) A Defect Study and Classification of Brown Diamonds with Non-Deformation-Related Color. *Minerals*, 10, 914; doi: 10.3390/min10100914

Hanson R.C., Jones L.H. (1981) Infrared and Raman studies of pressure effects on the vibrational modes of solid $CO_2$, *J. Phys. Chem.* 75, 1102-1112; doi: 10.1063/1.442183

Hunter J.D. (2007) Matplotlib: A 2D Graphics Environment. *Comput. Sci. Eng.*, 9 (3), 90–95.

Isenor M., Escribano R., Preson T.C., Signorell R. (2013) Predicting the infrared band profiles for $CO_2$ cloud particles on Mars, *Icarus,* 223(1), 591-601.

Khokhryakov A.F., Palyanov Y.N., Kupriyanov I.N., Nechaev D.V. (2016) Diamond crystallization in a $CO_2$-rich alkaline carbonate melt with a nitrogen additive, *J. Crystal Growth*, 449 (2016) 119–128

Kiflawi I., Mayer A.E., Spear P.M., van Wyk J.A., Woods G.S. (1994) Infrared absorption by the single nitrogen and A defect centres in diamond. *Phil. Mag. B*, 69(6), 1141–1147.

Lai M.Y., Stachel T., Breeding C.M., Stern R.A. (2020) Yellow diamonds with colourless cores – evidence for episodic diamond growth beneath Chidliak and the Ekati Mine, Canada, *Mineralogy and Petrology*, 114, 91–103.

Lang A.R., Bulanova G.P., Fisher D., Furkert S., Sarua A. (2007) Defects in a mixed-habit Yakutian diamond: Studies by optical and cathodoluminescence microscopy, infrared absorption, Raman scattering and photoluminescence spectroscopy, *J. Crystal Growth*, 309, 170–180.

Lu R., Hofmeister A. M. (1995) Infrared Fundamentals and Phase Transitions in $CO_2$ up to 50 GPa. *Phys. Rev. B*, 52(6), 3985–3992. https://doi.org/10.1103/PhysRevB.52.3985.

Lucazeau, G. and Novak, A. (1973), Low temperature Raman spectra of $KHCO_3$ single crystal. J. Raman Spectrosc., 1: 573-586. https://doi.org/10.1002/jrs.1250010607.

Malogolovets V.G., Nikityuk N.I. (1978) Vibrational frequencies of diamond lattice induced by substitutional impurities. *Superhard Materials*, 5, 28-31.

Melton C.E., Salotti C.A., Giardini A.A. (1972) The observation of nitrogen, water, carbon dioxide, methane, and argon as impurities in natural diamonds. *Am. Mineral.*, 57, 1518-1523.

Melton C.E., Giardini A.A. (1974) The composition and significance of gas released from natural diamonds from Africa and Brazil. *Am. Mineral.*, 59, 775-782

Melton C.E., Giardini A.A. (1975) Experimental results and a theoretical interpretation of gaseous inclusions found in Arkansas natural diamonds. *Am. Mineral.*, 60, 413-417.

Melton C.E., Giardini A.A. (1981) The nature and significance of occluded fluids in three Indian diamonds. *Am. Mineral.*, 66, 746-750.

Olijnyk H., Jephcoat A.P. (1998) Vibrational Studies on $CO_2$ up to 40 GPa by Raman Spectroscopy at Room Temperature. *Phys. Rev. B*, 57(2), 879–888. https://doi.org/10.1103/PhysRevB.57.879

Osberg W.E., Hornig D.F. (1952) The vibrational spectra of molecules and complex ions in crystals. VI. Carbon dioxide. *J. Phys.Chem.*, 20, 1345-1347, doi: 10.1063/1.1700760

Palyanov Y.N., Kupriyanov I.N., Sokol A.G., Borzdov Y.M., Khokhryakov A.F. (2016) Effect of $CO_2$ on crystallization and properties of diamond from ultra-alkaline carbonate melt. *Lithos*, 265, 339-350, DOI: 10.1016/j.lithos.2016.05.021

Schrauder M., Navon O. (1993) Solid Carbon Dioxide in a Natural Diamond. *Nature*, 365(6441), 42–44. https://doi.org/10.1038/365042a0

Shiryaev A.A., Iakoubovskii K., Grambole D., Dubrovinskaia N.D. (2006) Spectroscopic study of defects and inclusions in bulk poly- and nanocrystalline diamond aggregates, *J. Phys.: Condens. Matter*, 18, L493-L501.

Shiryaev A.A., Wiedenbeck M., Hainschwang T. (2010) Oxygen in bulk monocrystalline diamonds and its correlations with nitrogen, *J. Phys.: Condens. Matter*, 22, 045801-06

Signorell R., Jetzki M., Kunzmann M., Ueberschaer R. (2006) Unraveling the Origin of Band Shapes in Infrared Spectra of $N_2O$-$^{12}CO_2$ and $^{12}CO_2$-$^{13}CO_2$ Ice Particles. *J. Phys. Chem. A*, 110(9), 2890–2897. https://doi.org/10.1021/jp053021u

Smith E.M., Kopylova M.G., Frezzotti M.L., Afanasiev V.P. (2014) Fluid inclusions in Ebelyakh diamonds: Evidence of $CO_2$ liberation in eclogite and the effect of $H_2O$ on diamond habit. *Lithos*, 216-217, 106-117.

Sverjensky D. A., Huang F. (2015) Diamond formation due to a pH drop during fluid–rock interactions. *Nature Comm.*, 6:8702, DOI: 10.1038/ncomms9702

Taylor, W. R., Jaques, A. L., Ridd, M. (1990) Nitrogen-defect aggregation characteristics of some Australasian diamonds: Time-temperature constraints on the source regions of pipe and alluvial diamonds. *Amer. Miner.,* 75, 1290-1310.

Tomilenko, A. A., Chepurov, A. I., Pal'yanov, Y. N., Pokhilenko, L. N., Shebanin, A. P. (1997) Volatile components in the upper mantle (from data on fluid inclusions). *Russ. Geol. Geophys.,* 38, 294–303.

Voznyak, D. K., Kvasnitsa, V. N., Kislyakova, T. Ya. (1992) Liquefied gases in natural diamond, *Geochemistry International*, 29(9), 107-112 (Translated from *Geokhimiya*, 29(2), 268-273).

Wang H., Zeuschner J., Eremets M., Troyan I., Willams J. (2016) Stable solid and aqueous $H_2CO_3$ from $CO_2$ and $H_2O$ at high pressure and high temperature. *Scientific Reports,* 6:19902. DOI: 10.1038/srep19902

Weiss Y., Kiflawi I., Navon O. (2013) The IR Absorption Spectrum of Water in Microinclusion-Bearing Diamonds, in: D. G. Pearson et al. (eds.), Proceedings of 10th International Kimberlite Conference, Volume 1, Special Issue of the Journal of the Geological Society of India.

Wojdyr M. (2010) Fityk: A General-Purpose Peak Fitting Program. *J. Appl. Crystallogr.* 43(5), 1126–1128. https://doi.org/10.1107/S0021889810030499

Zedgenizov D.A., Shiryaev A.A., Shatsky V.S., Kagi H. (2006) Water-related IR characteristics in natural fibrous diamonds. *Min Mag.*, 70, 219–229.

Zheng, W., Kaiser, R.I. (2007) On the formation of carbonic acid ($H_2CO_3$) in solar system ices, *Chem. Phys. Lett.,* 450, 55–60.